\documentclass[iop,apjl]{emulateapj}
\shortauthors{A. Lapi et al.}
\shorttitle{ICP: a Universal Pressure Profile?}

\slugcomment{Accepted by ApJL}

\begin{document}

\title{The Intracluster Plasma: a Universal Pressure Profile?}
\author{A. Lapi\altaffilmark{1,2}, A. Cavaliere\altaffilmark{1,3}, R. Fusco-Femiano\altaffilmark{4}}
\affil{$^1$Dip. Fisica, Univ. `Tor Vergata', Via Ricerca Scientifica 1, 00133
Roma, Italy.}\affil{$^2$SISSA, Via Bonomea 265, 34136 Trieste,
Italy.}\affil{$^3$INAF, Osservatorio Astronomico di Roma, via Frascati 33,
00040 Monteporzio, Italy}\affil{$^4$INAF-IASF, Via Fosso del Cavaliere, 00133
Roma, Italy.}

\begin{abstract}
The pressure profiles of the Intracluster Plasma in galaxy clusters show a
wide variance when observed in X rays at low redshifts $z\la 0.2$. We find
the profiles to follow two main patterns, featuring either a steep or a
shallow shape throughout both core and outskirts. We trace these shapes back
to a physical dichotomy of clusters into two classes, marked by either low
entropy (LE) or high entropy (HE) throughout. From X-ray observations and
Sunyaev-Zel'dovich stacked data at higher $0.2\la z\la 0.4$, we elicit
evidence of an increasing abundance of HEs relative to LEs. We propose this
to constitute a systematic trend toward high $z$; specifically, we predict
the pressure profiles to converge into a truly universal HE-like template for
$z\ga 0.5$. We submit our physical templates and converging trend for further
observational tests, in view of the current and upcoming measurements of
individual, stacked, and integrated Sunyaev-Zel'dovich signals.
\end{abstract}


\section{Introduction}

\setcounter{footnote}{0}

A keen interest is focusing on the radial pressure profiles $p(r)$ that
prevail in the hot intracluster plasma (ICP) filling up the galaxy clusters.
Specifically, it is debated the degree of their `universality' among rich
clusters, on the following grounds.

On the upside, the profile of the ICP thermal pressure $p\equiv n\, k_BT/\mu$
(with the mean molecular weight $\mu\approx 0.59$ for cosmic abundances) is
expected to score off the components: temperature $T(r)$ and number density
$n(r)\equiv \rho(r)/m_p$, on account of its prompt equilibration at sound
speed in the absence of forcing stresses. The equilibrium gradient is simply
linked by
\begin{equation}
{\mathrm{d}p\over \mathrm{d}r} = - {G M(<r)\over r^2}\, \rho
\end{equation}
to the gravitational force from the dark matter (DM) distribution $M(<r)$ on
the ICP mass density $\rho$.

On the downside, one may worry that the equilibrium pressure profiles might
be affected by the \emph{complexities} elicited for redshifts $z\la 0.2$ in
the X-ray observations. These probe $n$ from the surface brightness
$S_X\propto n^2\, T^{1/2}$ emitted by the ICP via thermal bremsstrahlung,
while the temperatures $k_B T$ in the keV range are measured from
spectroscopy.

In fact, in the cluster cores for $r\la 0.2\, R_{500}$ \footnote{We adopt the
flat cosmology with matter density parameter $\Omega_M=0.3$, Hubble constant
$H_0=70$ km s$^{-1}$ Mpc$^{-1}$, and mass variance $\sigma_8=0.8$ (Komatsu et
al. 2011). The virial radius reads $R\approx R_{100}\approx
4\,R_{200}/3\approx 2\, R_{500}$ in terms of the radii encircling average
overdensities $100$, $200$ and $500$ over the critical density, and
takes on values around $2$ Mpc for rich clusters.} the radiation observed in
X rays is to erode the ICP thermal content on the cooling timescale
$t_c\approx 30\, (k_B T/{\rm keV})^{1/2}$ $(n/10^{-3}~{\rm cm}^{-3})^{-1}$
Gyr (e.g., White \& Rees 1978; Voit \& Bryan 2001). The process tends to
speed up as $n$ raises and $T$ lowers, to the effect of steepening the inner
pressure gradient after Eq.~(1). A catastrophic runaway is conceivably offset
by the energy fed back from self-regulated AGN activities (see discussions by
Cavaliere et al. 2002; Lapi et al. 2003, 2005; Ciotti \& Ostriker 2007;
Churazov 2010), or is even forestalled by strong energy injections from deep
mergers (see McCarthy et al. 2007; Markevitch \& Vikhlinin 2007).

In the cluster outskirts for $r\ga R_{500}$, a pressure jump is forced by
shocks driven by the supersonic inflow of cold external gas. Thus the
gravitational infall energy is mainly thermalized at the virial boundary (see
Lapi et al. 2005, 2010; Voit et al. 2005); meanwhile, part of it drives outer
subsonic turbulence contributing to the equilibrium so as to require lower
thermal pressures (see Lau et al. 2009; Cavaliere et al. 2011a).

Bypassing such complexities, a `universal' fitting formula for the pressure
profiles has been proposed by Nagai et al. (2007) to interpret the outcomes
of hydrodynamical simulations of relaxed clusters (see also Battaglia et al.
2011), and applied by Arnaud et al. (2010) to render with empirically
adjusted parameters the X-ray data out to $R_{500}$ for $z\la 0.2$. Actually,
these analyses led to recognize an average profile along with a considerable
\emph{variance}.

On the other hand, the pressure profiles can be \emph{directly} probed
with the thermal Sunyaev-Zel'dovich effect (1980; SZ) that occurs as CMB
photons are inverse Compton scattered by the hot ICP electrons, and change
the radiation temperature $T_{\rm cmb}\approx 2.73$ K by an amount $\Delta
T=g_\nu\, y\, T_{\rm cmb}\sim -0.5$ mK. This provides a linear, intrinsically
$z$-independent probe of the thermal electron pressure $p_e=\mu\,
p/\mu_e\approx 0.52\, p$ (with the mean molecular weight per electron
$\mu_e\approx 1.15$), since its strength is given by
the Comptonization parameter $y\equiv (\sigma_T/m_e c^2)\,
\int{\mathrm{d}\ell}\, p_e(r)\sim 10^{-4}$ integrated along the l.o.s. The
spectral factor $g_\nu$ approaches the value $-2$ at low frequencies (see
Rephaeli 1995); its positive signature for $\nu>217$ GHz offers a powerful
cross-check for the SZ nature of the signals.

The SZ observations now start to probe the radial profiles in nearby
individual clusters, and in more distant stacked samples (\textsl{SPT}
collaboration, Plagge et al. 2010; \textsl{WMAP} collaboration, Komatsu et
al. 2011; \textsl{Planck} collaboration, Aghanim et al. 2011a, 2011b). They
are also addressing the cluster contribution to the CMB power spectrum at
multipoles $\ell\ga 2000$ (see Lueker et al. 2010; Dunkley et al. 2011;
Reichardt et al. 2011). Extensive data at higher resolutions and
sensitivities are expected from current and upcoming instrumentation, and
eventually from \textsl{ALMA} (see Birkinshaw \& Lancaster 2007). All such
actively pursued observations call for a reliable template to interpret and
assess their astrophysical and cosmological import.

Here we take advantage of the effective formalism provided by the Supermodel
(SM; Cavaliere et al. 2009, Fusco-Femiano et al. 2009, see
\texttt{http://people.sissa.it/$\sim$lapi/ Supermodel/}) to show that the
complex ICP thermal states still allow a neat \emph{physical} description of
the spherically-averaged pressure profiles. In response to the intriguing
challenge posed by Arnaud et al. (2010), we find \emph{two} basic shapes to
span the observed variance in the pressure profiles for low $z\la 0.2$, and
predict that they are to converge into a closely \emph{universal} one for
high $z\ga 0.5$.

\section{A physical approach to pressure profiles}

In singling out the ICP disposition and evolution we base on the updated
paradigm for the hierarchical formation of the containing DM halos (e.g.,
Zhao et al. 2003; Genel et al. 2010; Wang et al. 2011). The paradigm
comprises an early collapse punctuated by major mergers, building up the core
over a few crossing times; this occurs at redshifts $z_f\approx 1.5-0.5$
weakly depending on the mass $M\sim 10^{14}-10^{15}\, M_\odot$. Slow,
dwindling inflows of external matter follow over several Gyrs, and accrue the
outskirts out to the current virial $R$ in closely stationary conditions
described by the Jeans equation. In the process, the core scale radius
$r_{-2}$ stays put while $R$ expands, and the `concentration' parameter
$c\equiv R/r_{-2}$ correspondingly grows.

The ICP forms from intergalactic gas with pressure $\la 10^{-3}$ eV cm$^{-3}$
(see Ryu et al. 2008; Nicastro et al. 2010) that -- along with the DM --
inflows at supersonic Mach numbers $\mathcal{M}\ga 10$. The gas is
shock-heated at about $R$, and its pressure jumps by factors $4\,
\mathcal{M}^2\ga 500$ up to the ICP values $p_R\ga 1$ eV cm$^{-3}$. Inward of
$R$, the pressure rises to balance the DM gravitational pull as described by
Eq.~(1), with the following \emph{universal} features: the rise will be
monotonic within a smooth potential well; the gradient will vanish at the
center when the gravitational force does, as implied by the Jeans equilibrium
and found in simulations and real data (see Lapi \& Cavaliere 2009a, 2009b;
Navarro et al. 2010; Newman et al. 2011); the rise terminates with a finite
central value, proportional to the thermal energy density.

Within such universal constraints, a full description of the pressure profile
is keyed to the specific `entropy' (adiabat) run $k(r)\equiv k_BT/n^{2/3}$
embodying the ICP thermal state. In fact, using $n\propto (p/k)^{3/5}$
Eq.~(1) solves to yield
\begin{equation}
p(r)=\left[p_R^{2/5}+{2\over 5}\,\int_r^R{\mathrm d}x~{m_p\,G\, M(<x)\over
x^2\, k^{3/5}(x)}~\right]^{5/2}~,
\end{equation}
in the context of the SM (see Lapi et al. 2005; Cavaliere et al. 2009).
For the DM mass distribution $M(<r)$ we use our $\alpha-$profiles;
these solve the Jeans equations under physical boundary conditions, and agree
well with $N-$body simulations (see Lapi \& Cavaliere 2009a).

\begin{figure*}
\epsscale{1.0}\plotone{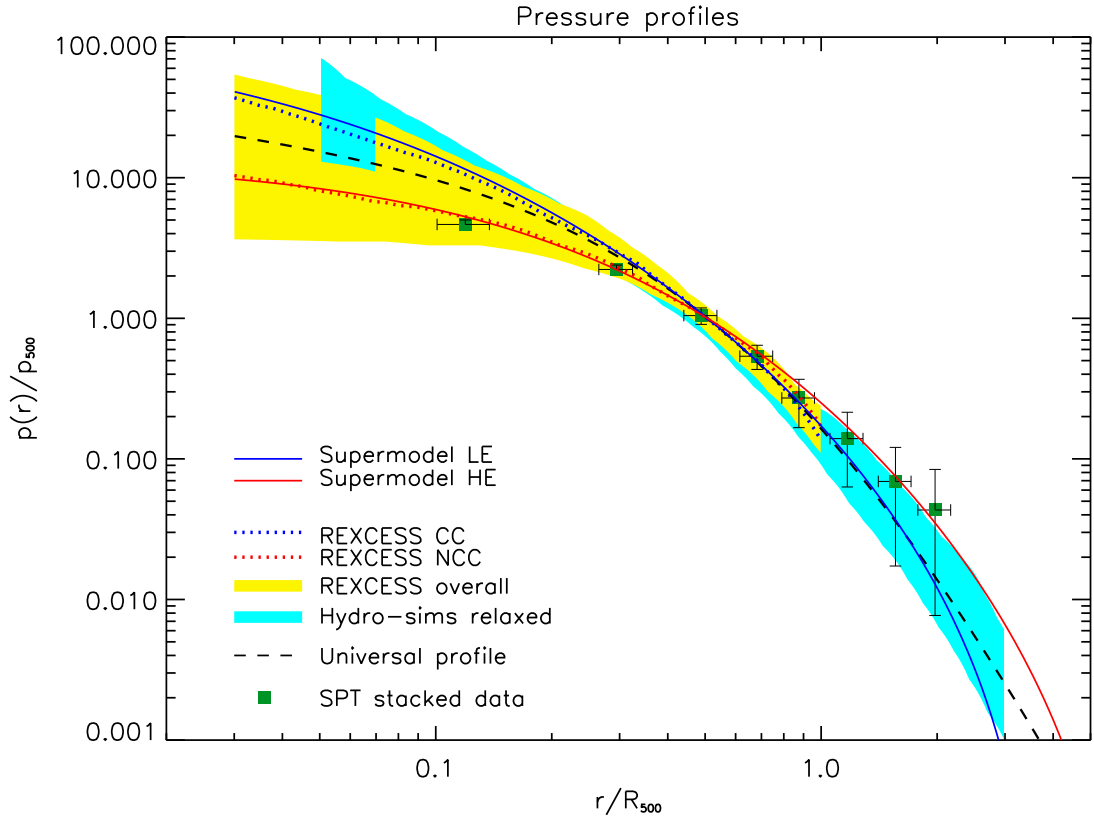}\caption{Profiles of ICP pressure normalized to
$p_{500}$, see \S~3 for details. The yellow shaded area illustrates the
region covered by the low redshift ($z\la 0.2$) clusters of the
\textsl{REXCESS} X-ray sample; the dotted blue and red lines refer to the
average profiles for the subsamples of cool-core and non-cool-core clusters,
separately. The cyan shaded area illustrates the coverage by hydrodynamical
simulations of relaxed clusters. The dashed line represents the joint fit to
the observational and virtual data with the `universal' pressure profile. The
green squares represent stacked SZ observations of higher redshift ($0.2\la
z\la0.4$) clusters with the \textsl{SPT}. Our SM templates for HE and LE
clusters are illustrated by the red and blue solid lines, respectively.}
\end{figure*}

The key role is played by the entropy run $k(r)$ over scales larger than some
$10^2$ kpc, for which physical insight is provided by the above picture of
cluster formation. The latter indicates the basic \emph{shape} (see also Voit
2005)
\begin{equation}
k(r)=k_c+k_R\,(r/R)^a~;
\end{equation}
this involves the boundary scales $R$ and $k_R$, and features two
\emph{intrinsic} parameters: the central level $k_c$, and the average outer
slope $a$. These are evaluated as follows.

In the cluster core $r\la 0.2\, R_{500}$ a level $k_c\sim 10^2$ keV cm$^2$ is
set by the early collapse; this results from densities $n\sim 10^{-3}$
cm$^{-3}$ compressed by the standard contrast factor $200$ over the average
background's, and from temperatures impulsively raised to values $k_B T\sim G
M(<r)/10\,r\sim$ a few keVs. Thereafter, radiative cooling competes with
energy injections from AGNs or mergers to the effect of stabilizing or even
raising the time-integrated $k_c$ at levels that gather around $10^1$ or
$10^2$ keV cm$^2$ (see Cavagnolo et al. 2009; Pratt et al. 2010; Hudson et
al. 2010). From Eqs.~(1) and (2) the central pressure is seen to follow the
basic scalings $p_c\propto k_c^{-5/8}$ and
$\mathrm{d}p^{2/5}/\mathrm{d}r\propto k_c^{-3/5}$.

At the other end $r\ga R_{500}$, an entropy ramp rises with slope $a\la 1.1$,
originated by the continuously shocked infall and the progressive
stratification of the accreted shells during the slow outskirts growth (see
Tozzi \& Norman 2001; Lapi et al. 2005). The ramp ends up at the
bounday $R$ with the value $k_R\ga 10^3$ keV cm$^2$ set by strong shocks (see
Cavaliere et al. 2011b). From Eqs.~(1) and (2) the outer pressure profile
follows the basic scaling $p(r)\simeq p_R\, (r/R)^{-5+2\,a}$, see Cavaliere
et al. (2011b). We stress that $p(r)$ will decrease monotonically outwards
even when the temperature $T(r)\propto p^{2/5}(r)\, k^{3/5}(r)$ features a
middle peak at $r\approx 0.2\, R_{500}$ due to the entropy rising steeply
from a low $k_c$; the peak is the defining mark of cool-core clusters (see
Molendi \& Pizzolato 2001; Leccardi et al. 2010).

When the X-ray data on brightness and temperature of clusters at $z\la 0.2$ are analyzed
with the SM, a direct \emph{correlation} between $k_c$ and $a$ emerges (see
Cavaliere et al. 2011b). This implies that clusters can be parted into two
main classes: LE or HE, featuring low or high entropies, respectively,
\emph{throughout} cores and outskirts. The LEs (e.g., A2204, A1795) are
marked by low central entropies $k_c\sim 10^1$ keV cm$^2$ along with shallow
entropy slopes $a\la 0.7$; the HEs (e.g., A1656, A399) are marked by higher
central values $k_c\ga 10^2$ keV cm$^2$ along with steeper slopes $a\approx
1$.

Then from the above inner and outer scalings ${\rm }dp^{2/5}/{\rm d}r\propto
k_c^{-3/5}$ and $p\propto r^{-5+2\,a}$, we find the pressure profiles to
\emph{differ} from HEs to LEs, with the former featuring quite
\emph{shallower} gradients both in the core and in the outskirts. Our picture
is substantiated by the data presented in Fig.~1.

\begin{figure*}
\epsscale{1.1}\plottwo{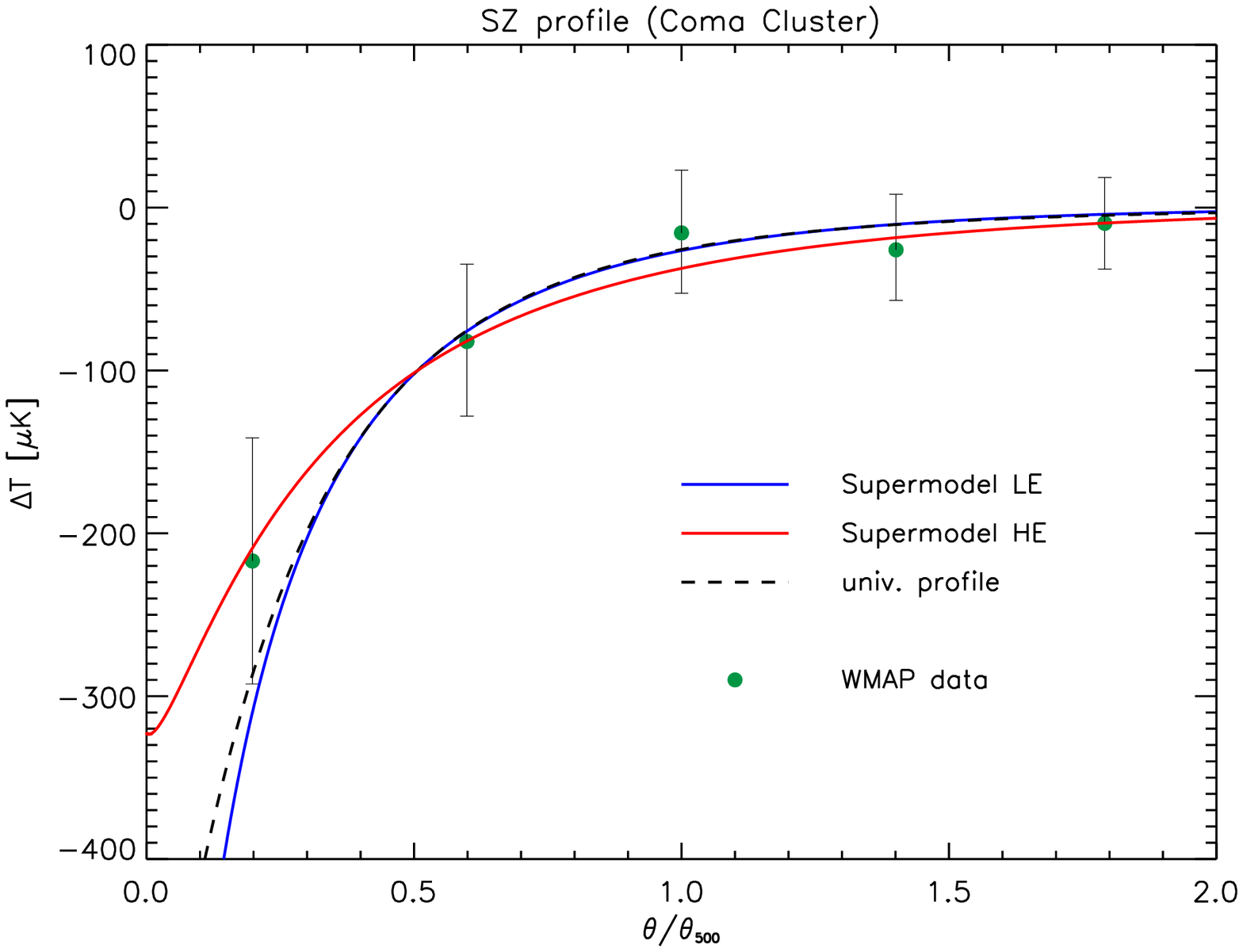}{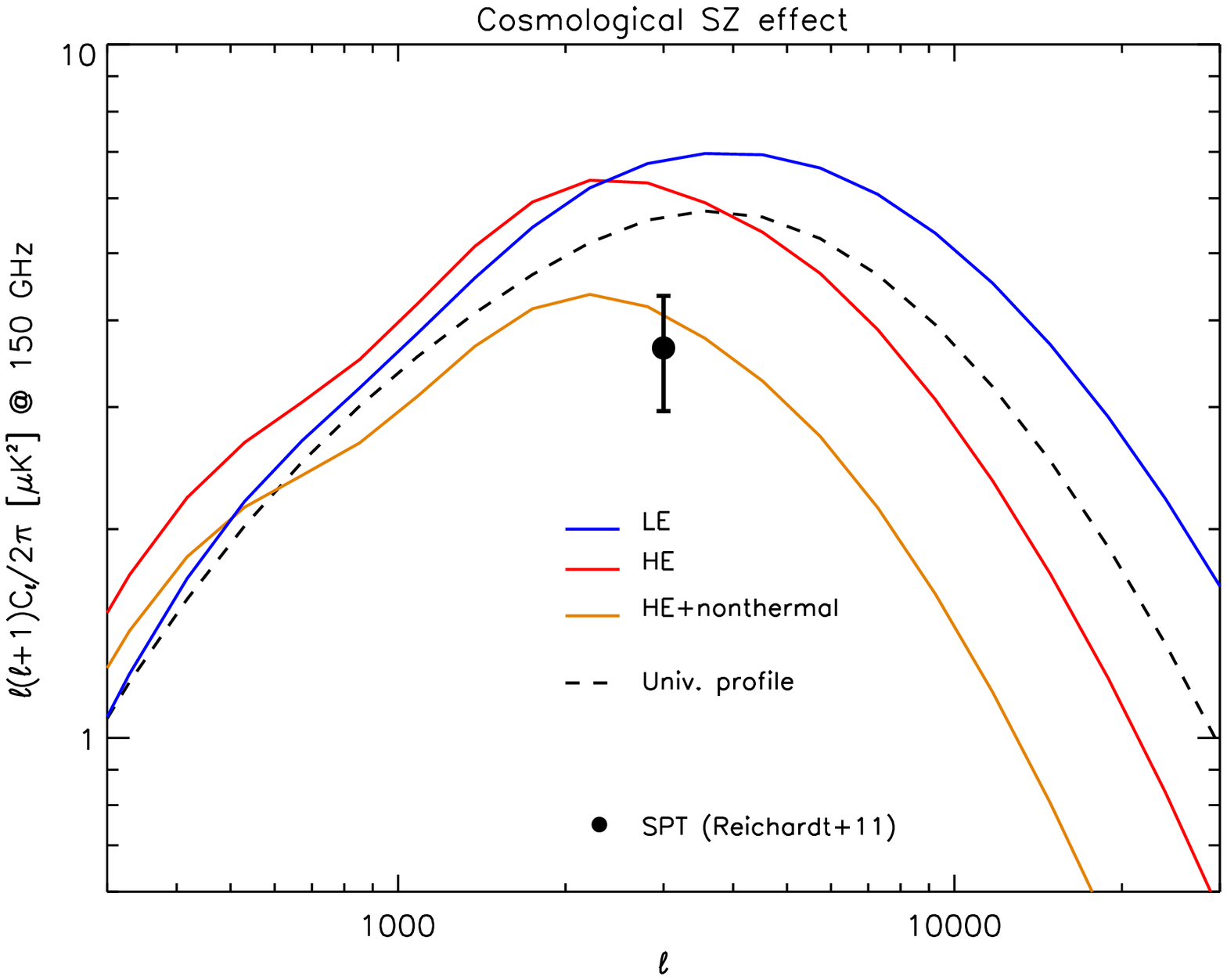}\caption{\emph{Left panel}: Profile
of the SZ signal from the Coma Cluster, in terms of Rayleigh-Jeans
temperature decrement $\Delta T$. The green filled dots illustrate the
\textsl{WMAP} data. The black dashed line refers the `universal' pressure
profile. The red and blue solid lines illustrate our SM outcomes for HE and
LE clusters. \emph{Right panel}: Cosmological thermal SZ effect at $150$ GHz.
The black dot with error bars illustrates the current constraints set
with the \textsl{SPT}. The dashed line refers to the `universal' pressure
profile. The outcomes from our SM templates are illustrated by the solid
lines: blue for LEs, red for HEs, and orange for HEs with a
nonthermal component increasing with $z$ (see \S~3).}
\end{figure*}

\section{Probing pressure with X rays and SZ effect}

In Fig.~1 we compare the pressure profiles computed with our SM to data from
X-ray and SZ observations, and from numerical simulations. In the plot, the
radial scale is normalized to $R_{500}$, while the pressure is normalized to
the standard value $p_{500}\approx 1.8\, h_z^{8/3}$ $(M_{500}/5\times
10^{14}\, M_{\odot})^{2/3}$ eV cm$^{-3}$ in terms of the Hubble parameter
$h_z\equiv H(z)/H_0$ (e.g., Ettori et al. 2004; Arnaud et al. 2010).

Our pressure templates provided by Eq.~(2) for HE and LE clusters are
illustrated by the red and blue solid lines; these are computed from Eqs.~(2)
and (3) with the parameter values discussed in \S~2. Specifically, for
typical HEs we adopt $k_c=100$ keV cm$^2$, $k_R=3\times 10^3$ keV cm$^2$ and
$a=1.1$, while for LEs we adopt $k_c=10$ keV cm$^2$, $k_R=10^3$ keV cm$^2$,
and $a=0.7$; for both we set $R=2$ Mpc. These values are consistent with the
outcomes from detailed SM fits to X-ray observations of nearby clusters
carried out by Fusco-Femiano et al. (2009) and Cavaliere et al. (2011b);
note, however, that fitting the centrally flat profiles of HEs (e.g.,
A2256) requires the level $k_c$ to extend out to $r_f\sim 10^2$ kpc, bearing
the imprint of a recent merger.

The yellow shaded area illustrates the region covered by the low redshift
($z\la 0.2$) clusters of the \textsl{REXCESS} X-ray sample analyzed by Arnaud
et al. (2010); the dotted blue and red lines refer to their average profiles
for the subsamples of cool-core and non-cool-core clusters, respectively. The
cyan shaded area illustrates the region covered by hydrodynamical simulations
of relaxed clusters (Borgani et al. 2004; Nagai et al. 2007; Piffaretti \&
Valdarnini 2008; Battaglia et al. 2011).

The dashed line represents the joint fit by Arnaud et al. (2010) to the
observational and virtual data in terms of their `universal' pressure
profile. For $r\la R_{500}$ the X-ray data show that such a profile yields
only an average but incomplete description. In fact, the partial averages
over the cool-core and non-cool-core subsamples deviate upward and downward
by a large amount exceeding their internal variance; thus a \emph{bimodal}
description constitutes both a closer and a more effective representation.
This is just what is provided by the above SM templates for HE or LE
clusters, which in the core recover the non-cool-core or cool-core behaviors.
Moreover, for $r\ga R_{500}$ where only scarce X-ray data are available, the
SM template for LE clusters agrees well with the results of
hydro$-$simulations of relaxed clusters. In the way of a \emph{prediction},
beyond $R_{500}$ we expect for HEs considerably higher pressure profiles
relative to LEs, as represented in Fig.~1.

In addition, our picture envisages for decreasing $z$ an ICP evolution from
HE to LE states. In fact, over lifetimes of several Gyrs elapsed from the
formation $z_f\approx 1$ to the observation redshift $z\la 0.2$ the outer
slope $a$ will flatten out from values $a\approx 1.1$ towards values $a\la
0.5$, and correspondingly the boundary level $k_R$ decreases; such a trend is
enhanced as $z$ approaches $0$, when the cosmic timescale lengthens
considerably. This is because the entropy production by weakened virial
shocks is \emph{reduced} as inflows peter out, especially in the accelerated
cosmology (see Lapi et al. 2010). Meanwhile, cooling \emph{erodes} the
central entropy $k_c$ over comparable times $t_c\approx 5\, (k_c/100\,
\mathrm{keV~cm}^2)^{1.2}$ Gyr after Eq.~(2), and undergoes an accelerated
drop toward the `attractor' level $k_c\sim 10$ keV cm$^2$ set by competition
with AGN feedback.

On the other hand, the evolution of the cluster DM halo is marked by the
growth of the `concentration' $c\approx 3.5\,h_{z_f}/h_z$ from $z_f$ to $z$
(see Zhao et al. 2003; Prada et al. 2011). Since $a$ and $k_c$ decrease
together, we expect them both to \emph{anticorrelate} with $c$. In fact, such
correlations $a$, $k_c$ vs. $c^{-1}$ have been quantitatively elicited with
SM analyses of high-quality X-ray data concerning several clusters at $z\la
0.2$ (see Cavaliere et al. 2011b).

As $z$ increases, our evolutionary trend envisages the HE/LE ratio to grow
towards $1$; in other words, we expect all pressure profiles to converge
toward a truly \emph{universal} template provided by the HE shape. Our
picture is supported by comparison of the local data with stacked SZ
observations of redshifts $0.2\la z\la 0.4$ clusters; the pressure profiles
from the \textsl{SPT} stacked data (Plagge et al. 2010) are represented in
Fig.~1 with the green squares. Although the uncertainties are still
considerable in the outskirts, a departure from the `universal' profile
stands out, and the trend toward an HE-like template clearly emerges. The
same trend is emerging from the analysis of a stacked cluster sample observed
with $\textsl{WMAP}$ (Komatsu et al. 2011). A similar trend is suggested by
the sample of clusters detected by $\textsl{Planck}$ for redshift $0.3\la
z\la 0.5$, and followed up in X rays with $\textsl{XMM-Newton}$ (Aghanim et
al. 2011b). Independent evidence is provided by the dearth of strong
cool-cores found in X rays by Santos et al. (2010) at high-redshifts.

The next step to test the evolutionary trend from LEs to HEs will involve
observing the SZ profile from individual clusters over an extended range of
redshifts. This is still challenging with the present instrumentation even at
$z\la 0.2$, but will become feasible out to high $z\ga 0.5$ with
new-generation instruments up to \textsl{ALMA} (see
\texttt{http://www.almaobservatory.org/}). To illustrate the current status,
in Fig.~2 we report the recent data on the SZ profile for the Coma Cluster at
$z\approx 0.023$ with \textsl{WMAP} (Komatsu et al. 2011). We also report the
`universal' pressure profile (black dashed line), and the Supermodel
templates for HE and LE clusters (red and blue solid lines). These data allow
to recognize the HE nature of the Coma cluster, still more prominent in X
rays. On the other hand, the SZ data will become competitive once resolutions
better than $1^{\prime}$ will be attained.

Another testbed for our evolutionary picture will be provided by the power
spectrum of the unresolved SZ effect integrated over redshift and over the
evolving cluster mass distribution including groups (e.g., Shaw et al. 2010;
Efstathiou \& Migliaccio 2011). In Fig.~3 we illustrate the outcome from the
`universal' profile of Arnaud et al. (2010) and the SM templates for LE and
HE clusters, compared with the constraints at $\ell\sim 3000$ set by current
observations with the \textsl{SPT} (Reichardt et al. 2011). The
latter are converging to indicate that the integrated SZ effect is to be both
dominated by HEs at the relevant $z\ga 0.5$, and reduced by the presence of a
nonthermal contribution to the inner ICP equilibrium.

We find consistency of our HE template with the data when the
standard scaling $p_{500}\propto h_z^{8/3}$ is decreased by a factor
$h_z^{-1/2}$; the latter renders how the inner nonthermal component mildly
increases with $z$, as expected for the early HE equilibrium punctuated by
major mergers with strong turbulent wakes (see \S~2 and Cavaliere et al.
2011a). Additional constraints on the detailed shape of the SZ effect will
require high sensitivities at resolutions $\ell\ga 5000$ with good control of
the systematics, an effort actively pursued by the SZ community.

\bigskip
\bigskip

\section{Discussion and conclusions}

As the observational uncertainties in the SZ measurements are shrinking below
the theoretical ones, we have aimed at eliciting truly universal features in
the pressure profiles of the Intracluster Plasma (ICP). These are described
in terms of our Supermodel, are tested at low $z$ against X-ray observations
of brightness and temperature, and are proposed for probing at higher $z$
with the linear, $z$-independent SZ effect.

At low $z\la 0.2$ we identify from X-ray data two main pressure patterns,
each featuring related shapes at the center and in the outskirts. These are
interpreted in terms of two cluster classes: HEs with high entropy
\emph{throughout}, i.e., high central levels $k_c$ along with a steep outer
rise with slope $a$, leading to shallow pressure profiles; conversely for
LEs.

For $z\ga 0.2$ we expect the onset of an evolutionary \emph{trend} in cosmic
time from HE to LE, comprising fast erosion of central entropy by radiative
cooling, along with reduced outer entropy production by decreasing inflows
(see Cavaliere et al. 2011b). Accordingly, the ratio HE/LE is to increase
toward high $z$, implying for $z\ga 0.5$ convergence of the pressure profiles
toward a truly \emph{universal}, HE-like template.

Few intermediate instances are expected; these may occur when cooling is
forestalled by an exceedingly high initial level $k_c>10^2$ keV cm$^2$, while
the outer entropy ramp is independently flattening (e.g., in A1689 and A2218
as discussed by Cavaliere et al. 2011b). On the other hand, Eq.~(2) implies
our radial pressure profiles to be robust against overall asphericity and
localized hot/cold imprints from recent mergers, even when the latter cause
wiggles in temperature and marked flatness in the central brightness as often
found from detailed fits to X-ray data in HEs (e.g., A2256 as discussed by
Fusco-Femiano et al. 2009). This strengthens the case for universality of our
asymptotic HE-like pressure template.

We have traced evidence for our evolutionary trend developing in the pressure
profiles through stacked SZ signals due to $0.2\la z\la 0.4$ clusters (see
Fig.~1). At higher $z$, further evidence will be difficult to pinpoint from X
rays alone, given their bias toward high central brightnesses proper to LEs;
rather, it will be provided again by SZ signals from individual or stacked
clusters. In parallel, such an evidence will be tested with the integrated
contribution from unresolved SZ signals to the CMB anisotropies at multipoles
$\ell\ga 5000$ (see Fig.~2). All such observations require high-sensitivity
data at resolutions below $1^{\prime}$, that are pursued with current instrumentation
and will eventually culminate with \textsl{ALMA}.

\begin{acknowledgements}
Work supported in part by MIUR, ASI and INAF. We thank M. Massardi, M.
Migliaccio, and D. Nagai for helpful discussions, and our referee for useful
comments. A.L. thanks SISSA for warm hospitality.
\end{acknowledgements}

\end{document}